# Possible Ground-State Structure of $Au_{26}$: a Highly Symmetric Tube-Like Cage

Wei Fa and Jinming Dong*

*Group of Computational Condensed Matter Physics, National Laboratory of Solid State Microstructures, and Department of Physics, Nanjing University, Nanjing 210093, China*

**Abstract.** A surprising stability of the tube-like $Au_N$ ($N$ = 26-28) has been shown using the scalar relativistic all-electron density functional theory calculations, forming another powerful candidate for the lowest-energy $Au_N$ competing with those previously suggested space-filled structures. Unlike the icosahedral "golden" fullerene $Au_{32}$, these tube-like gold clusters may be closely related to the synthesized single-wall gold nanotubes (SWGNT). The ground-state $Au_{26}$ has a hollow tube-like structure constructed from the (6, 0) SWGNT, yielding a high-symmetry $D_{6d}$ cage, based upon which the most stable $Au_{27}$ and $Au_{28}$ can be obtained by capped one and two atoms, respectively.

## I. Introduction

In the past decades, gold clusters have attracted great interests of both experimental and theoretical researchers due to their fundamental importance and tremendous potential applications in catalysis, biology, and nanoscale devices.[1-3] Recently, a novel tetrahedral structure of $Au_{20}$ has been identified by the photoelectron spectroscopy (PES) in juxtaposition with the relativistic density functional theory (DFT) calculations,[4] which is found to be a fragment of the face-centered cubic lattice of bulk gold with a small structural relaxation. Such a remarkable discovery of the $T_d$ $Au_{20}$ challenges the previous conclusions on the space-filled compact structures of gold clusters in the medium-sized range,[5-10] and stimulates further searches for their real ground-state structures by including more possible structural competitors. More recently, a DFT study predicted the existence of the highly stable icosahedron $Au_{32}$ cage, constructed from the carbon fullerene $C_{60}$ as a template and so referred as "golden" fullerene.[11] The $Au_{32}$ fullerene cage satisfies the $2(N+1)^2$ aromatic rule[12] and has an extremely large energy gap between the highest occupied molecular orbital (HOMO) and the lowest unoccupied molecular orbital (LUMO) up to at least 1.5 eV, showing its chemical stability and its potential ability to assemble into molecular crystals. Subsequently, a similar $I_h$ $Au_{42}$, born out of the $C_{80}$ fullerene, was also reported theoretically to be competitive energetically with respect to its space-filled isomers.[13] The cage-like gold clusters would be of both fundamental and practical interests because of their high surface area (all the atoms are on the cluster surface) and hollow cavity to accommodate other atoms or molecules.

Despite these theoretical advances, there is still no experimental evidence for the golden fullerene cages until now. The PES of the anionic $Au_{32}^-$ measured at finite temperatures revealed that it has a closed-shell electronic structure with a relatively small energy gap of 0.30 eV, differing distinctly from the electronic characteristics of the $I_h$ $Au_{32}$.[14, 15] On the other hand, the multi-wall and single-wall gold nanotube (SWGNT) were synthesized,[16, 17] offering another possibility to form a hollow tube-like $Au_N$ by closing a segment of the SWGNT with appropriate caps, which would be a powerful candidate, competing in energy with other structures, e.g., the amorphous, bulk-fragment, and fullerenic ones. Based upon the scalar relativistic all-electron DFT calculations, we have already shown the existence of such a kind of $Au_N$,[18] since the tube-like $Au_{24}$ and $Au_{26}$ are more stable than their previously suggested "lowest-energy" isomers with space-filled compact structures.[8-10] However, their tube surfaces are distorted anisotropically and heavily, leading to the low-symmetric tube-like structures only. From the aesthetical point of view, it is naturally expected to find a more symmetric $Au_N$ with tube-like configurations. Also, it is interesting to seek other non-space-filled gold clusters as potential true ground states for the bigger sizes. Therefore, further computational efforts are still needed to explore the energy landscapes of these medium-sized gold clusters.

Here, we will report our extensive studies on about 70 samples of Au$_N$ ($N$ = 26-28) using DFT calculations, in which we pay particular attention to the tube-like structures constructed from the SWGNTs with different helicities and diameters (from 4.5 Å to 8.3 Å). The following section introduces the DFT procedures used for our calculations. The obtained results and discussions are given in section III. Some concluding remarks are offered in section IV.

**II. Computational Approach**

Our calculations were carried out using the spin-polarized DFT at the level of a generalized gradient approach (GGA) using a Perdew-Wang exchange-correlation functional.[19] The molecular DFT DMol[3] package was employed,[20] in which a double-numerical basis set together with polarization functions was chosen to describe the electronic wave functions. In order to find the lowest-energy structures more efficiently, the calculated stratagem we adopted was divided into two steps. Firstly, we used the relativistic semi-core pseudopotential (RECP)[21] to do geometrical optimizations from as many as possible initial candidates, including not only available structures in the literatures[8-10, 18] but also new ones obtained from an unbiased global search with the guided simulated annealing[22] to an empirical interaction potential[23] and a short segment of a SWGNT closed by appropriate ends. Secondly, at least fifteen lowest energetic samples determined from the first step were chosen to do further structural relaxations within the scalar relativistic all-electron DFT schemes. The cluster geometry was optimized by the Broyden - Fletcher - Goldfarb - Shanno algorithm[24] without symmetry constraints until the total energy was converged to $10^{-6}$ eV in the self-consistent loop and the force on each atom was less than 5 meV/Å. The frequency check has been performed to validate the stability of the ground-state Au$_N$ we found. The vibrational spectra were calculated in the harmonic approximation using the finite displacement technique to obtain the force constant matrix. Positive and negative displacements with the magnitude of 0.005 Å were used in order to obtain more accurate central finite differences. The IR intensities were obtained from the derivative of the dipole moment.

The reliability of our method was tested by benchmark calculations on the gold atom and the dimer. The ionization potential of 9.68 eV and the electron affinity of 2.21 eV were obtained for the gold atom, agreeing with the corresponding experimental data of 9.22 eV[25] and 2.31 eV,[26] respectively. The calulated equilibrium distance of Au$_2$ is 2.49 Å that is in good agreement with the experimental value of 2.47 Å.[27, 28] We also note that the calculated binding energy of 2.45 eV and vibration frequency of 184 cm$^{-1}$ for the Au$_2$ are in well consistent with the experimental data of 2.28 ± 0.10 eV and 191 cm$^{-1}$, respectively.

**III. Results and Discussions**

As illustrated in Fig. 1, the lowest-energy Au$_{26}$ can be viewed as a short segment of the (6, 0) SWGNT capped by two atoms at both ends, yielding a highly symmetric tube-like cage with $D_{6d}$ symmetry. It has two different kinds of sites, i.e., 12 five-coordinated sites and 14 six-coordinated sites. To check stability of this new tube-like Au$_{26}$, we have compared it with other isomers, including those reported previously as the "global minima".[8-10, 18] The isomer (b) is amorphous, which can be regarded essentially as a much distorted cage with one atom inside, but its energy is 0.439 eV higher than the tube-like $D_{6d}$ Au$_{26}$. The isomer (c) is also tube-like, obtained from a distorted piece of the (7, 0) SWGNT capped with a triangle on each side,[18] which is separated from the ground-state structure with $D_{6d}$ symmetry by 0.444 eV. By analyzing the structural characteristics of the isomer (c), it is found that its distribution of interatomic distances shows a sharp peak at the nearest-neighbor distance, separated by a wider gap from a more continuous distribution because the isomer (c) has a low-symmetric $C_2$ structure in contrast to the high-symmetric $D_{6d}$ Au$_{26}$ with well-separated values of the interatomic distances. There are other tube-like isomers lying at the higher energy, such as the isomer (d) based upon the (8, 4) SWGNT (see Table 1 and Fig 1). It is particularly necessary to explore the space-filled isomers, characterized by a broader nearest-neighbor distance distribution with an almost zero or much smaller gap in the whole distance range, since they were often considered as the most stable structures of the medium-sized Au$_N$. A space-filled compact structure (e) of Au$_{26}$ is separated from the tube-like $D_{6d}$ one by a large energy gap of 0.862 eV, which was predicted as the "ground state" of the Sutton-Chen potential.[8] Another space-filled structure with an approximate $D_{3h}$ symmetry (g), suggested by N. T. Wilson et al. using a Murrell-Mottram potential,[9] is also less stable than the tube-like Au$_{26}$ by 1.437 eV. Our structure search has found many other space-filled isomers close to each other in energy, forming an almost continuous energy distribution, which may constitute a broad funnel of energy landscape to trap the previous searches for the lowest-energy gold clusters. In addition, the planar and bulk-fragment structures have been considered, in which a planar isomer with $C_s$ symmetry (f) is the most stable but still 1.412 eV higher than the ground state found here. We also started from a gold fullerene taking the $D_{3h}$ $C_{26}$ as a template and found it is unstable. Finally, we notice that the HOMO-LUMO gap of the lowest-energy tube-like $Au_{26}$ is 0.842 eV, which is appreciably larger than those of other isomers, showing also its high stability. Therefore, this highly symmetric tube-like $Au_{26}$ we found provides another powerful structure competitor with other isomers, e.g., amorphous, bulk-fragment, and fullerenic cage-like, which may lead to a deeper insight in the structural properties of $Au_N$.

The frequency check has been carried out to make sure that the obtained structure is a true minimum rather than the saddle point on the potential energy surface. No imaginary frequencies were found for the lowest-energy tube-like $Au_{26}$, indicating its stability. Six of the calculated $3N$ frequencies are zero in magnitudes, corresponding to rotational and translational degrees of freedom. The remaining $3N$-6 frequencies are the vibrational normal modes, from which the vibration density of states (VDOS) are obtained by a convolution with a Lorentzian broadening of 2 cm$^{-1}$. As shown in Fig. 2, the normal frequencies of the tube-like $D_{6d}$ $Au_{26}$ locate in the range of 23-185 cm$^{-1}$. Only a portion of its vibration modes are infrared-active due to its high symmetry, and as a results, its far-infrared (FIR) spectrum exhibits some well-resolved peaks. We hope that the infrared resonance-enhanced multiple photon dissociation measurements can be used to acquire the FIR absorption spectrum of $Au_{26}$, as those done for vanadium and niobium clusters,[29] so that the existence of our tube-like $D_{6d}$ $Au_{26}$ can be validated. In addition, we also notice that the tube-like $D_{6d}$ $Au_{26}$ has a radial breathing-like mode at 119 cm$^{-1}$, characteristic of the tube-like or cage structures, which is similar to radial breaking modes found in the carbon nanotubes, and so may be tested by future Raman spectra.

Since the tube-like structure may be another powerful competitor of the ground-state $Au_N$, it would not manifest itself only in the tube-like $Au_{24}$[18] and $Au_{26}$. As expected, the $Au_{27}$ and $Au_{28}$ with tube-like structures are also found to be stable (see Fig. 3), which can be regarded as having one and two capped atoms located on the hollow $D_{6d}$ $Au_{26}$, respectively. By comparing the energies of the tube-like $Au_{27}$ and $Au_{28}$ to their other isomers, we have found that the energy differences favor our tube-like configurations as the ground states. For example, the tube-like $Au_{27}$ (27-a) is 0.82 eV (27-b) and 1.76 eV (27-c) lower than those of the space-filled structures with $C_s$ symmetry, respectively. [8, 9] Though we cannot guarantee absolutely that the "ground-state" structures of Au$_N$ ($N$ = 26-28) we found are their true global minima, we think that at least they are quite stable, and by far the lowest-energy structures of Au$_N$ ($N$ = 26-28), known by the comparisons with their possible isomers, including the previously suggested ones.

The underlying mechanism for the stability of the tube-like gold clusters may be related to the relativistic effect (RE), which can be demonstrated by comparing the non-relativistic results with the scalar-relativistic ones, and also the results of other coinage metal clusters with those of Au$_N$. For example, the non-relativistic tube-like Au$_{26}$ with $D_{6d}$ symmetry is 0.53 eV higher in energy than its space-filled $C_s$ isomer.[8] On the other hand, the tube-like Ag$_{26}$ and Cu$_{26}$ with much less RE are not stable compared to their compact structures. The obvious RE in Au$_N$ manifests itself in the shorter inter-atomic distances, leading to a stronger $s$-$d$ hybridization and $d$-$d$ interaction, which makes Au$_N$ favor the tube-like structure with less but more stronger bonds. On the other hand, considering the existence of the carbon nanotubes and boron-nitrogen nanotubes too, in which there is no RE, the stability of the tube-like Au$_N$ and SWGNTs is an interesting and important open problem, deserving further experimental and theoretical studies.

## IV. Conclusions

In summary, we have found that the Au$_{26}$ prefers a highly symmetric tube-like structure without core atoms, which is constructed from a segment of (6, 0) SWGNT capped by two atoms on both sides, showing a close relationship to the synthesized SWGNT. And so, such a hollow tube-like structure may exist more frequently in the larger Au$_N$, competing in energy with other possible structures, i.e., the space-filled ones, the bulk fragments, and the golden fullerenes. As expected, the ground-state Au$_{27}$ and Au$_{28}$ are also found to be tube-like, which is derived from the tube-like $D_{6d}$ Au$_{26}$ with one and two capped atoms, respectively. The ultimate driving force behind them remains unknown, but may be related to the RE. These tube-like Au$_N$ with larger surface area and hollow inner space may be important for fundamental chemistry researches, and in future, have a wide range of applications in catalysis, biology, and nanotechnology, etc.

**ACKNOWLEDGMENT.** This work was supported by the Natural Science Foundation of China under Grants No. 90503012 and No. A040108, and also by the state key program of China through Grant No. 2004CB619004. All computations were performed on the SCI Origin-3800 and 2000 supercomputers.

TABLE I. The scalar relativistic all-electron DFT results on the energy ($E$), the HOMO-LUMO gap ($E_g$), the average bond length ($\bar{d}$), and the average coordination number ($\bar{Z}$) of the ground-state and selected low-lying $Au_{26}$. All energies are relative to that of the tube-like $D_{6d}$ $Au_{26}$. The references to the previously reported structures are also listed.

| Structure | $E$ (eV) | $E_g$ (eV) | $\bar{d}$ (Å) | $\bar{Z}$ | Note |
|---|---|---|---|---|---|
| (a) Tube-like cage based on the (6, 0) SWGNT | 0 | 0.842 | 2.707 | 5.538 | - |
| (b) No symmetry isomer with one core atom | 0.439 | 0.410 | 2.742 | 6.00 | - |
| (c) Tube-like cage based on the (7, 0) SWGNT | 0.444 | 0.596 | 2.714 | 5.538 | Ref. 18 |
| (d) Tube-like cage based on the (8, 4) SWGNT | 0.541 | 0.326 | 2.715 | 5.538 | - |
| (e) Compact isomer with two core atoms | 0.862 | 0.373 | 2.780 | 6.462 | Refs. 8, 10 |
| (f) Planar structure | 1.412 | 0.288 | 2.665 | 4.615 | - |
| (g) Compact isomer with three core atoms | 1.437 | 0.177 | 2.790 | 6.385 | Ref. 9 |

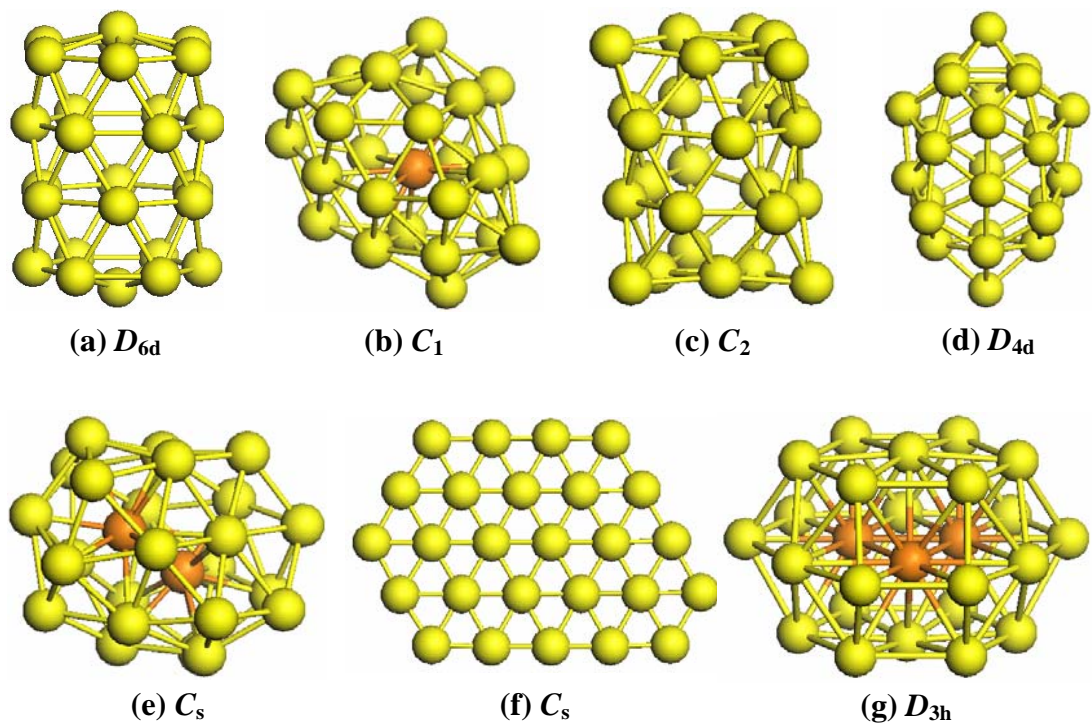

**(a)** $D_{6d}$  **(b)** $C_1$  **(c)** $C_2$  **(d)** $D_{4d}$

**(e)** $C_s$  **(f)** $C_s$  **(g)** $D_{3h}$

FIG. 1. (Color online) Optimized ground state (a) and selected low-lying isomers (b-g) of $Au_{26}$.

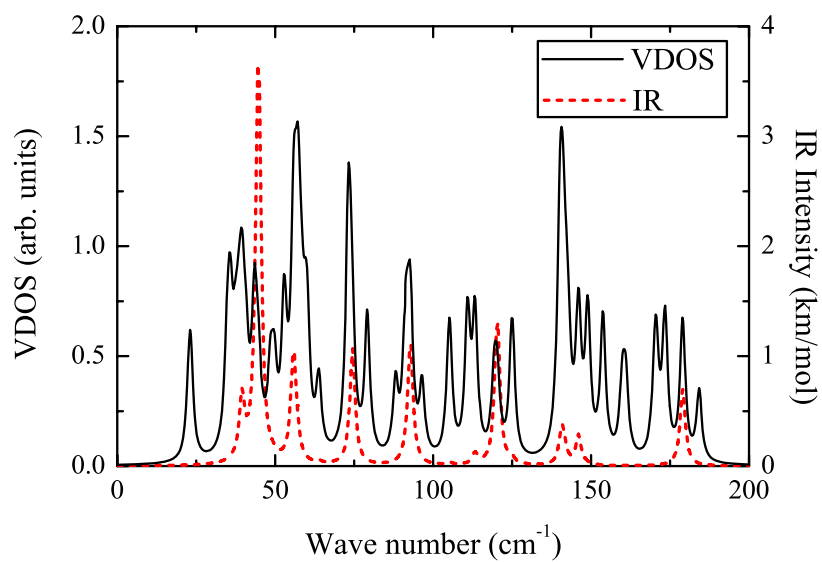

FIG. 2. (Color online) The VDOS and corresponding FIR absorption spectrum of the lowest-energy tube-like $Au_{26}$, shown by the solid and dashed line, respectively.

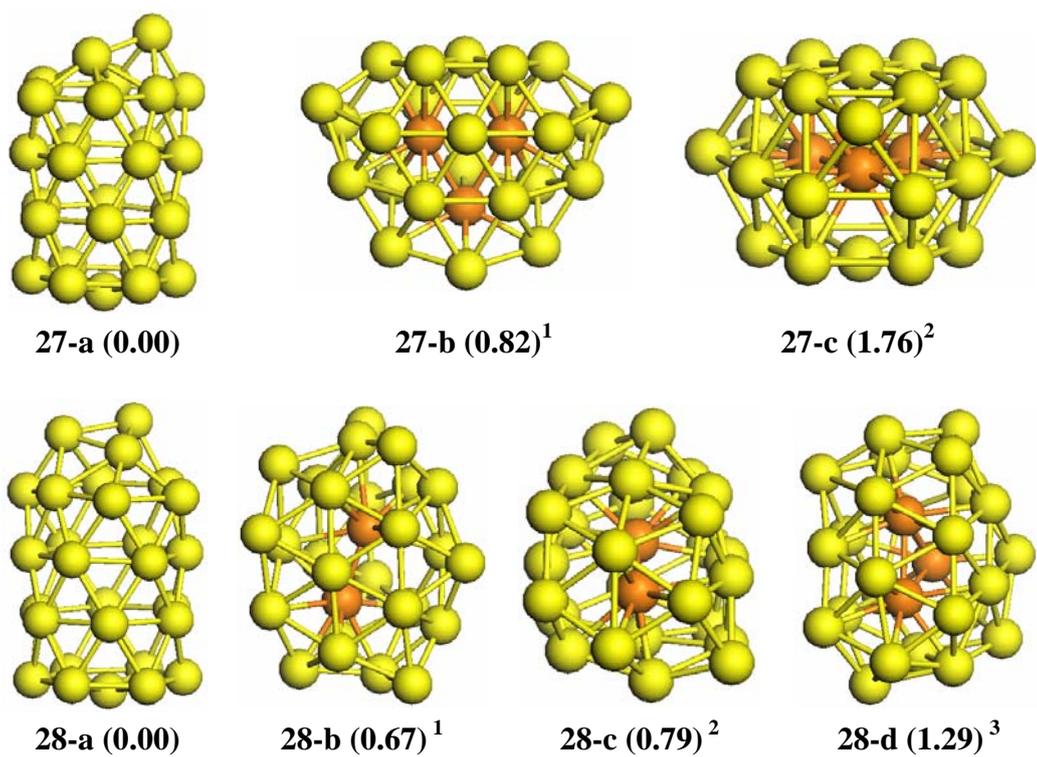

FIG. 3. (Color online) Structures of $Au_{27}$ and $Au_{28}$. The values in the parentheses are the relative energies (in unit of eV) to those of the tube-like structures. [1]See Ref. 8; [2]See Ref. 9; [3]See Ref. 10.